      [1999/12/01 v1.4c Il Nuovo Cimento]
\documentclass{cimento}

\usepackage{graphicx}  
\title{The unexpected clustering of the optical afterglow luminosities}
\author{M. Nardini\from{ins:oab}\from{ins:sissa}\thanks{e-mail:
             nardini@sissa.it.}\ETC, 
G. Ghisellini\from{ins:oab},
G. Ghirlanda\from{ins:oab},
F. Tavecchio\from{ins:oab}\\
C. Firmani \from{ins:2}
        \atque
D. Lazzati\from{ins:col}}
\instlist{\inst{ins:oab} Osservatorio Astronomico di Brera, via
Bianchi 46, 23807 Merate, Italy  
\inst{ins:sissa} 
SISSA--ISAS, via Beirut 2-4, 34014 Trieste, Italy. 
  \inst{ins:2} Instituto de Astronomia UNAM, AP 70-264,04510 M\'exico,
  DF,  M\'exico
\inst{ins:col} JILA, University of Colorado, Boulder, CO 80309-0440,USA}  
\begin{document}

\maketitle

\begin{abstract}
We studied the behaviour of the optical afterglow lightcurves of a sample
of 24 Gamma--Ray Bursts (GRBs) with known redshift and
published estimates of the optical extinction in the source frame,
detected before the SWIFT satellite launch. 
We found an unexpected clustering of the optical luminosities at 12
hours in the source frame. 
The distribution of the optical luminosities is narrower than  the
distribution of X--ray luminosities at the same time.  
Few (3) bursts stand apart from the main optical distribution, 
being fainter by a factor of about 15. 
We also analysed the optical luminosities of the SWIFT burst with known
redshift finding that the luminosity distribution is similar to 
the pre SWIFT GRBs one, even if they have a different mean redshift. 
These results can suggest the existence of a family of intrinsically 
optically under--luminous dark GRBs.  
    
\end{abstract}

\section{Pre--SWIFT optical luminosity lightcurves}

In order to study the common features of the long GRB
optical afterglow emission, we analysed the $R$--band lightcurves
of a sample of long GRBs with known redshift detected before the
launch of the SWIFT satellite. 
For better comparing the different lightcurves we only considered 
those events with a published estimate of the host galaxy dust absorption. 
We found 24 GRBs satisfying these requirements.
\\
The observed lightcurves, despite showing a similar mean 
temporal behaviour with a decay well described by powerlaw in the form
$F(\nu_R)\propto \nu^{-\alpha}$, are spread by several orders of
magnitude. 
At 12 hours after the trigger, the observed
fluxes at the mean wavelength $\nu_R=6400$\AA~cover a range between
$-3\leq\log{F(\nu_R)}[mJy]\leq 1$ with a log--normal distribution
described by a mean value $\mu=-1.03$ and a dispersion $\sigma\approx 0.48$.
We converted the observed fluxes into the intrinsic monochromatic
luminosities in the host galaxy frame, k--correcting them using the published
estimates of the optical spectral index $\beta_o$ and considering the
host galaxy dust extinction. 
A similar procedure was already done by \cite{ref:gb05} for the 
X--ray afterglows. 
We found an unexpected clustering of the optical luminosities,
calculated at the same rest frame time (12 hours after trigger).
The corresponding distribution is much narrower than the distribution
of the observed fluxes.
Most of the considered GRBs (21 on 24) show a similar luminosity. 
The log--normal distribution has a mean value $\mu=30.65$ [erg s$^{-1}$ Hz$^{-1}$] 
and a dispersion $\sigma \approx 0.28$. 
The remaining 3 differ from the majority by more than 4 $\sigma$ being
under--luminous by a factor of about 15. 
No event was found in the luminosity range 
$29.7<\log{L(\nu_R)}$ [erg s$^{-1}$ Hz$^{-1}$]$<30.2$. 
The X--ray afterglow luminosities do not cluster as much.  
\\
We also studied the optical to X--ray spectral energy
distribution of these GRBs. Most of them are consistent with the
synchrotron emission process, while 2 GRBs show a possible Inverse
Compton component in the X--ray band.  For all but two GRBs, the
cooling frequency appears between optical and the X--ray bands. In GRB011121
$\nu_c<\nu_{opt}$ and GRB020813 $\nu_c>\nu_{10keV}$.  

\begin{figure}
\center
\includegraphics[scale=0.239]{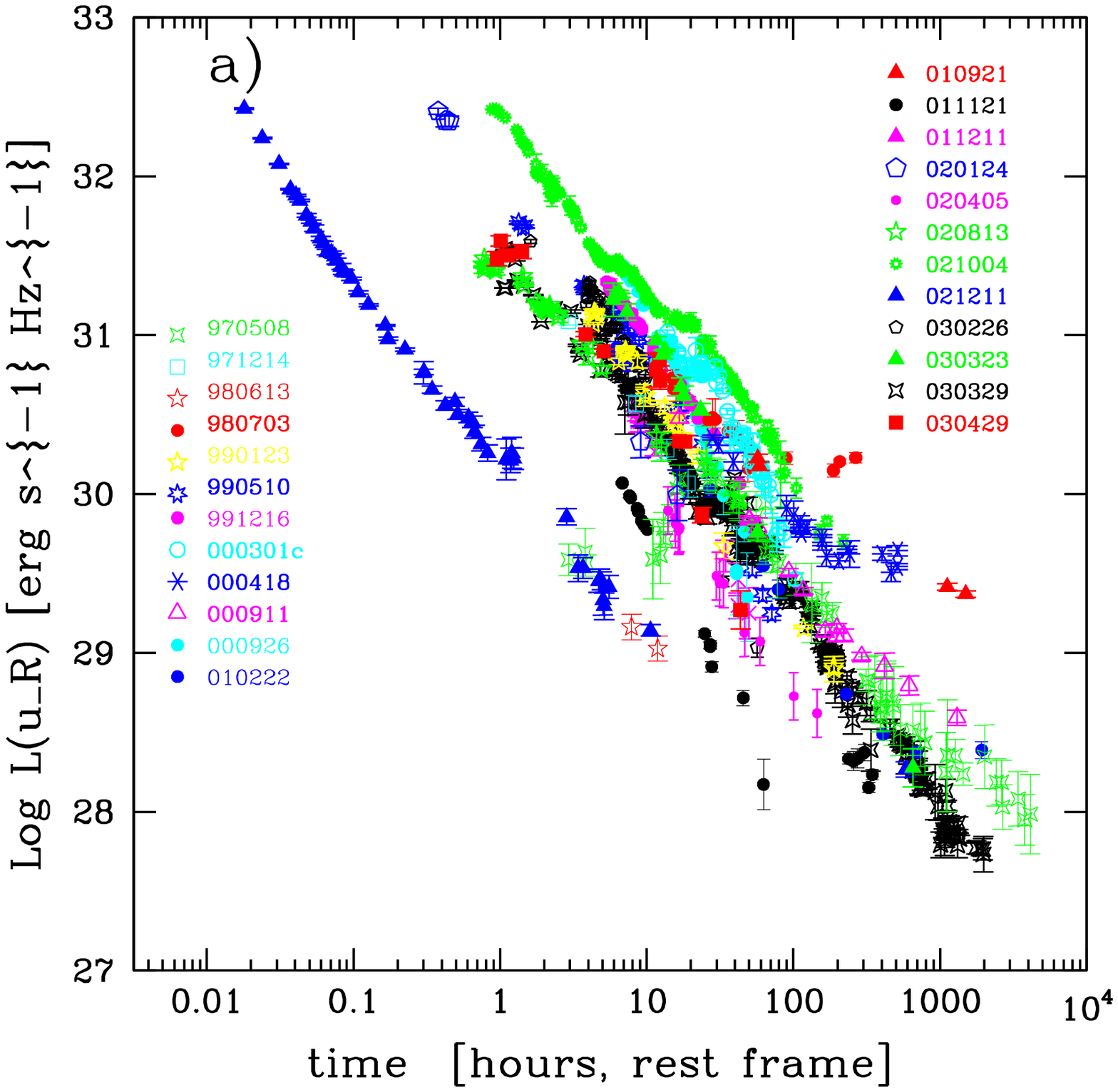}  
\includegraphics[scale=0.239]{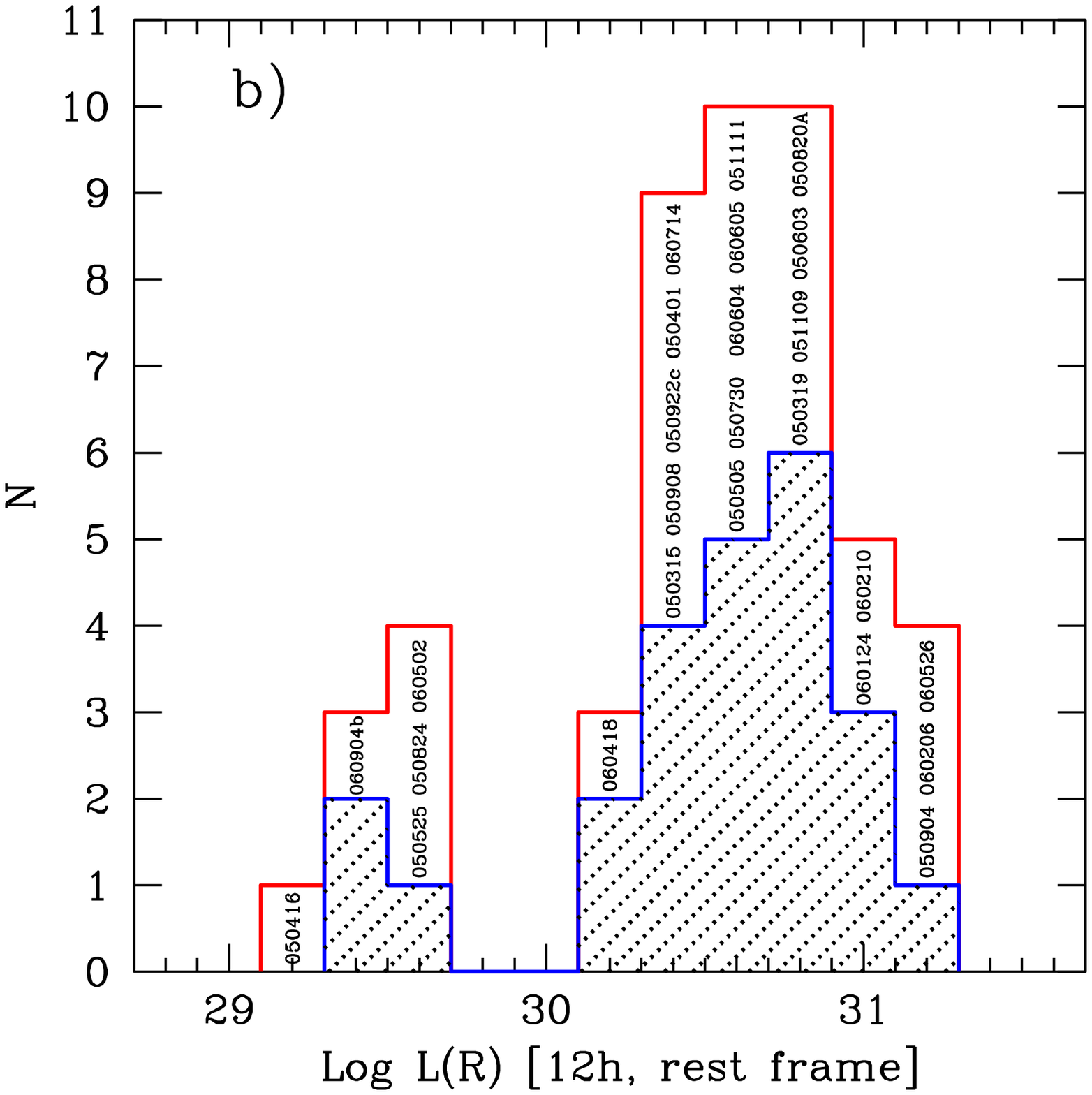}  
\label{swift}  
\caption{a) Monochromatic luminosity lightcurves of the 24 pre--SWIFT
GRBs. \cite{ref:io} b) Histogram of the SWIFT GRBs $\log{L(\nu_R)}$ at
12 h (host frame), superposed to the pre--SWIFT one.} 
\end{figure}

\section{Update with the GRBs observed by SWIFT}

Since its launch, the SWIFT satellite detected more than 150 long
GRBs. 40 bursts have a spectroscopic
redshift determination; 25 of them have enough photometry to
determine $L(\nu_R)$ at 12 hours in the host frame and 8 of them have
a published estimate of the host galaxy dust absorption.
In Fig. \ref{swift} we show the luminosities of these SWIFT GRBs 
together with the 24 pre--SWIFT GRBs. 
Despite the different mean redshift of the long GRBs detected by
SWIFT, the addition of these bursts confirms both the clustering of
the bright optical afterglow luminosities and the hint of bimodality
in the distribution. This confirmation opens the question if the
optically under--luminous bursts are the tip of the iceberg of an
optically dark GRBs family.

\end{document}